# Glass Hardness: Predicting Composition and Load Effects via Symbolic Reasoning-Informed Machine Learning

Sajid Mannan[1], Mohd Zaki[1], Suresh Bishnoi[1], Daniel R. Cassar[2], Jeanini Jiusti[3], Julio Cesar Ferreira Faria[4], Johan F. S. Christensen[5], Nitya Nand Gosvami[6], Morten M. Smedskjaer[5], Edgar Dutra Zanotto[4, *], N. M. Anoop Krishnan[1,7,*]

[1]Department of Civil Engineering, Indian Institute of Technology Delhi, Hauz Khas, New Delhi, 110016, India

[2]Ilum School of Science, Brazilian Center for Research in Energy and Materials (CNPEM), Campinas, Brazil

[3]Laboratory of study and development of nuclear waste conditioning matrices, French Alternative Energies and Atomic Energy Commission, France

[4]Department of Materials Engineering - Federal University of São Carlos, Brazil

[5]Department of Chemistry and Bioscience, Aalborg University, 9220 Aalborg, Denmark

[6]Department of Materials Science and Engineering, Indian Institute of Technology Delhi, Hauz Khas, New Delhi, 110016, India

[7]Yardi School of Artificial Intelligence, Indian Institute of Technology Delhi, Hauz Khas, New Delhi, 110016, India

*Corresponding authors: krishnan@iitd.ac.in (NMAK), dedz@ufscar.edu (EDZ)

**Abstract**

Glass hardness varies in a non-linear fashion with the chemical composition and applied load, a phenomenon known as the indentation size effect (ISE), which is challenging to predict quantitatively. Here, using a *curated* dataset of over ~3,000 inorganic glasses from the literature comprising the composition, indentation load, and hardness, we develop machine learning (ML) models to predict the composition and load dependence of Vickers hardness. Interestingly, when tested on new glass compositions unseen during the training, the standard data-driven ML model failed to capture the ISE. To address this gap, we combined an empirical expression (Bernhardt's law) to describe the ISE with ML to develop a framework that incorporates the symbolic law representing the domain reasoning in ML, namely *Symbolic Reasoning-Informed ML Procedure* (SRIMP). We show that the resulting SRIMP outperforms the data-driven ML model in predicting the ISE. Finally, we interpret the SRIMP model to understand the contribution of the glass network formers and modifiers toward composition and load-dependent (ISE) and load-independent hardness. The deconvolution of the hardness into load-dependent and load-independent terms paves the way toward a holistic understanding of composition and ISE in glasses, enabling the accelerated discovery of new glass compositions with targeted hardness.

1. **Introduction**

Glasses are widely used in several applications involving interactions with other entities, namely, automobile windshields and other specialty windows, artificial gems, ballistic armors, smartphone and computer protective screens, biomedical implants, nuclear waste immobilization matrices, and artistic pieces [1,2]. The glass resistance to be damaged in such applications correlates with their hardness [3], which is typically determined by instrumented indentation experiments [4], [5–7]. However, the hardness values obtained from such experiments are *not* an intrinsic glass property; it also depends on other parameters, such as the loading procedure, geometry of the indenter, and environmental conditions [8,9]. Specifically, hardness monotonically decreases and saturates with increasing applied load—a phenomenon termed indentation size effect (ISE) [10–13]. This load-dependent plastic behavior of glass has been attributed to the stress concentration generated due to sharp contact loading, which causes localized structural changes in the glass network leading to permanent deformation [14]. The ISE prevents the comparison of hardness values obtained at different loading conditions. Predicting load-independent hardness values is thus crucial for comparing experiments conducted at varying loads.

Experimental studies have paved the way for several empirical laws capturing the load-dependence of glass hardness [10,15–18]. A widely used empirical law for predicting the ISE was provided by Bernhardt [19]. Such relations are based on experimental observations fitted to simple polynomial-like functions. Thus, these laws are based on human observation/intervention, which is hard-coded in symbolic expressions by testing on a large family of glasses, an approach known as symbolic reasoning [20]. Symbolic reasoning has found extensive success in materials science, where several observations have been coded as empirical rules, enabling the prediction of material properties. However, these empirical laws require the fitting parameters to be calibrated for each family of glasses and typically are not transferable from one family to another. These laws are useful but require multiple experiments for each glass family to obtain the fitting parameters.

Recently, researchers have been successful in predicting glass properties using machine learning (ML) techniques [21–29]. For instance, Ravinder et al. (2020) [21] developed neural network models for predicting the physical, mechanical, optical, and thermal properties of oxide glasses. Since these models were trained on a limited number of components (~30 components) and properties (eight properties), there was a need for models which could predict properties for a broader range of components encompassing oxides, halides, and sulfides, to name a few [30]. To overcome this limitation, researchers [22] developed tree-based machine learning models for predicting 25 properties of oxide glasses for a larger dataset. In another work [23], researchers predicted the hardness of oxide glasses as a function of their composition, testing load, and annealing temperature, which demonstrated improved property prediction and strengthened the fact that the glass hardness is a function of composition, testing, and processing parameters [23]. However, these models are purely data-driven and unable to incorporate any empirical or physics-based knowledge that severely limits the generalizability of the model to compositions outside the training space.

Moreover, interpreting the compositional dependence of properties is also required for designing novel glasses with targeted properties, a challenging task for black-box ML models. To this end, Alcobaça et al. (2020) [31] used explainable machine learning models to predict the glass transition temperature of oxide glasses. This work was followed by others on predicting and explaining the properties of glasses using large databases, employing *explainable* ML models and a game theory-based approach known as Shapely additive explanations (SHAP) [22,32,33]. Similar attempts have been made to predict the properties of chalcogenide glasses along with their explanation of compositional effect using SHAP analysis [29,34].

Some recent studies have attempted to predict the Vickers hardness of oxide glasses using either composition only or composition and testing conditions, such as applied load [35,36]. However, these models use a classical data-driven model to predict the hardness that does not incorporate domain knowledge to ensure better generalization. Hence, these models cannot capture the load dependency or ISE, as demonstrated later in the present work. ML has also been used to predict other mechanical properties based on instrumented indentation [37]. An alternate approach incorporating domain knowledge is called Physics-Informed ML [38], where known physical laws governing the system are included as part of the loss function. Such methods have succeeded in predicting glass properties, such as viscosity, which exhibits a temperature-dependent variation based on the MYEGA or VFT equations [39]. Physics-informed ML has also been used to predict glass structure, where the physical information is provided by an independent statistical model [40]. However, to the best of our knowledge, there has been no effort toward incorporating knowledge developed based on intuition and observations in ML, that is, combining *symbolic reasoning* and ML towards improved prediction and understanding of a glass response. Also, no work has used a manually *curated* composition–load–hardness dataset for training an ML model.

Here, we propose a symbolic reasoning-informed ML procedure (SRIMP) framework, combining Bernhardt's law and ML, to understand the composition and load dependence of glass hardness. The contribution of this work is twofold: (i) we present the first highly *curated* dataset on composition–load–hardness values of glasses, which is made available publicly, and (ii) we develop a *SRIMP framework* that allows the definition of a load-independent hardness, while also providing insights into the indentation size effect for a given glass composition. To this end, we manually collected a large dataset (~3000 entries) of composition, load, and corresponding hardness values for different inorganic glasses from the literature, which was used to train traditional data-driven ML models. We demonstrate that even the ML models trained with the load as an input feature fail to capture the ISE. In contrast, the SRIMP model predicts the hardness in an improved fashion and exhibits an excellent capability to predict the load-dependent hardness variation. Further, we consider indentation experiments on three different glass compositions [41] to evaluate the transferability of the approach to new experiments performed in-house, on which the model has no information. Finally, the SRIMP provides interpretability to the ML model, thereby allowing the definition of a load-independent hardness. We demonstrate that the load-independent hardness corresponds to the asymptotic hardness at high loads and allows the hardness prediction at any load without performing an experiment.

## 2. Methodology

*2.1. Dataset collection and preparation*

The datasets used in this work were manually collected from published literature on glass, that is, research papers focusing on Vickers' indentation experiments in oxide glasses. This dataset was normalized to ensure that the compositions added up to 100%.

All the glass compositions were represented in the elemental form, with the value of each element in a glass composition taken in atomic mole percentage. Also, the dataset was rounded off up to three decimal places, and then duplicate entries were merged by taking the mean of property values lying in the interval of ± 2.5 % of the mean of all corresponding property values after avoiding the outliers (if present, by considering two times standard deviation as a criterion). Additionally, we considered only those elements present in at least 30 or more glass compositions to ensure representative data in both training and test datasets. The above protocol resulted in hardness data for 3325 unique glasses with 52 features (51 different elements and load as the input features) for model training, testing, and further analysis. These data were used for developing the ML models using load and composition as the input.

However, for many of these compositions, the hardness values were available for only a single load value. Hardness values corresponding to at least three different loads are required to fit the parametric equation of ISE for any given glass composition. To this end, we further curated a smaller dataset of about approximately 150 glasses [15], [42]–[46], [46]–[48] from the entire dataset, for which the hardness values corresponding to three or more load values were available. These data were used to train the SRIMP model.

*2.2. Symbolic-reasoning informed machine learning procedure (SRIMP)*

The hardness, $H$, from indentation experiments was computed as

$$H = \frac{2P\sin(\theta/2)}{L_D^2} \quad , \tag{1}$$

where, $L_D$ is the diagonal length of the indent after the indenter is removed, $\theta$ is the tip angle of the indenter, and $P$ is the applied load. The ISE using Bernhardt's law [49] is

$$\frac{P}{L_D} = a_1 + a_2 L_D \qquad (2)$$

where $a_1$ and $a_2$ are fitting parameters. Substituting for $L_D$ from Eq. (1) in Eq. (2), we get

$$P = a_1 \sqrt{\frac{2P \sin(\theta/2)}{H}} + a_2 \frac{2P \sin(\theta/2)}{H} \qquad (3)$$

Solving this quadratic equation for hardness $H$, we get (see Supplementary materials for the detailed derivation)

$$H = \frac{C_1}{2P} + C_2 + \frac{\sqrt{C_1^2 + 4C_1 C_2 P}}{2P}, \qquad (4)$$

where $C_1 = 2a_1^2 \sin(\theta/2)$, and $C_2 = 2a_2 \sin(\theta/2)$.

Similarly, substituting for load $P$ in Eq. (1) from Eq. (2), we get

$$H = 2a_2 \sin(\theta/2) + \frac{2a_1 \sin(\theta/2)}{L_D} = H_\infty + \frac{a_{\text{ISE}}}{L_D}, \qquad (5)$$

where the parameters $H_\infty = 2a_2 \sin(\theta/2)$ and $a_{\text{ISE}} = 2a_1 \sin(\theta/2)$ correspond to the load-independent hardness and the extent of the ISE, respectively. Although Eqs. (4) and (5) are equivalent, one is obtained by eliminating $L_D$, and the other is obtained by eliminating $P$. Thus, Eq. (5) requires $L_D$, a quantity that can only be determined experimentally, while Eq. (4) allows the prediction of $H$ purely based on load.

Figure 1 shows a graphical representation of the methodology employed in SRIMP. We train the ML model to predict the $C_1$ and $C_2$ values associated with each glass composition by minimizing the loss function,

$$Loss = \frac{1}{n} \sum_{i=1}^{n} (H_{SRIML}^i - H_{actual}^i)^2 \qquad (6)$$

where $H_{actual}^i$ represents the measured hardness of a certain glass at a given load and $H_{SRIML}^i$ represents the predicted hardness using Eq. (4) for a total of $n$ data points. The loss is then backpropagated through the multilayer perceptron (MLP) to learn the optimal weights that minimize it. In this process, the ML model learns the function relating to the composition and the parameters $C_1$ and $C_2$. Further, by comparing Eqs. (4) and (5), the relationship between $C_1$ and $C_2$ with $a_{ISE}$ and $H_\infty$, respectively, can be obtained as $C_2 = H_\infty$ and $C_1 = \alpha_{ISE}^2/(2\sin(\theta/2))$. This approach directly allows interpreting any glass composition's ISE and the inherent hardness.

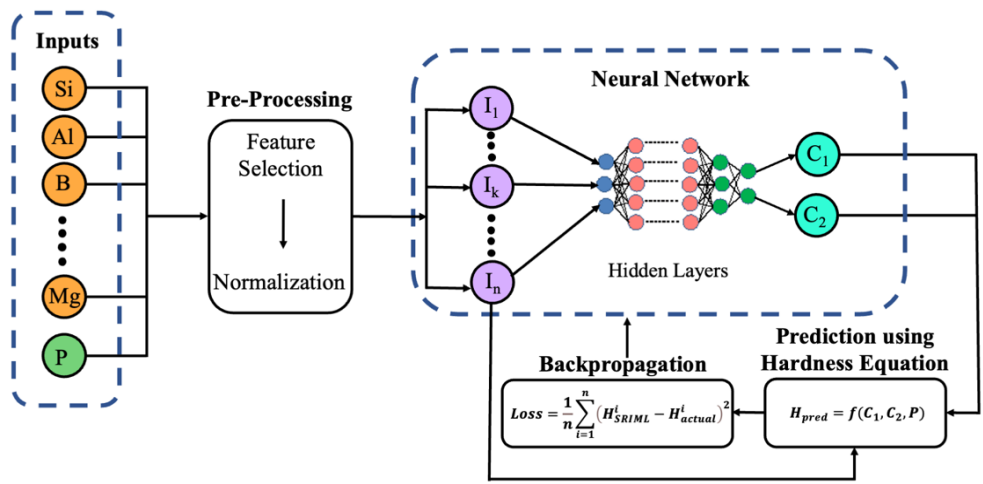

**Figure 1.** Workflow diagram for the SRIMP model.

*2.3. Model Training*

To train the ML models, the cleaned datasets (both larger and smaller) were divided into 80:20 proportions for the training and testing sets, respectively. Fig. 1 shows the schematic of the training scheme with load (P) as an input. All the features were taken as input parameters, and the hardness was the output parameter for the ML model training. The mean and standard deviation of the training set features were used to scale the data before training by computing the z-score. [50] The training set was subjected to 10-fold cross-validation to enable hyperparameter search for an optimal fit. Since machine learning models have different hyperparameters, they need to be chosen appropriately to achieve generalized model performance. In the case of NN, these hyperparameters are the number of hidden layers, number of neurons in each hidden layer, learning rate, weight decay, dropout probability, activation function, and optimizer. The hyperparameters of the ML models were optimized using the Bayesian optimization-based library, Optuna [51], employing a previously reported approach [22]. The model with the best $R^2$ score on the validation fold was chosen as the final model. Since the test data were kept hidden in the training step to avoid data leakage [52], they were used only to evaluate the performance of the best model.

The following python packages are used in this work: PyTorch, an open-source ML library [53], was used for developing a feed-forward neural network (NN). Batch normalization and dropout were used in the hidden layers to help in reducing the model overfitting to the data. NumPy, Pandas, and Matplotlib [54–56] were used for pre-processing and data visualization. Although we employed both tree-based and neural ML models, we obtained the best performance using an NN in terms of mean squared error and $R^2$ for both validation and training scores. The final hyperparameters used in the ML models are reproduced in Table 1. All the codes and data used for the present work are available at https://github.com/M3RG-IITD/SRIMP_Hardness

**Table 1.** Details of model hyperparameters.

| Hyperparameter | Model with Load | Model without Load |
|---|---|---|
| Epoch | 500 | 500 |
| Batch Size | 64 | 32 |
| Number of layers | 1 | 1 |
| Neurons | 92 | 97 |
| Dropout | 0.2 | 0.1 |
| Activation | LeakyReLU | LeakyReLU |
| Optimizer | Adam | Adam |
| Learning rate | 0.000832 | 0.000494 |
| Weight Decay | 0.000322 | 0.000416 |

*2.4. Shapley Additive Explanations*

To understand the model's output, we used the SHAP [57], an interpretable machine learning algorithm, to understand the influence of individual features on the output. It is a unified game theoretic approach where players and the game can be considered as the feature values of instances and the model output, respectively. Shapely values tell us how fairly the prediction is distributed among the features. Mathematically, Shapley values are calculated as [22]

$$\phi_i = \sum_{S \subseteq F\{i\}} \frac{|S|!\,(|F|-|S|-1)!\,|F|!}{|F|!} \left[ f_{S \cup \{i\}}(x_{S \cup \{i\}}) - f_s(x_s) \right] \qquad (7)$$

where F denotes the set of all features, S denotes a subset of features, $f_{S \cup \{i\}}$ denotes a model trained with the feature present, $f_{s\{i\}}$ denotes a model trained with the feature withheld, and $x_s$ denotes input features in subset S. By comparing the prediction of these models on the current input, SHAP determines the significance of a feature. The feature importance is high if the prediction error is high; otherwise, it is low. SHAP has introduced various model approximations such as kernel, Deep, Linear, and Tree-based explainers. Based on the task and the model, any explainer can be opted to calculate the SHAP values for each feature. The explainer exploits the model's internal complexity and collapses it to a low-order polynomial complexity [22].

Various plot types, such as bar, violin, river flow, and bee swarm plots, can be used to visualize the SHAP values. In this study, we use bar and violin plots to evaluate the ML model. Bar plots give the average impact of each feature on the model output without showing the directionality of effects, whether it is influencing positively or negatively. However, the violin plots represent the contribution of each feature towards the different model outputs as a function of the feature value. Thus, it is colored based on the feature value and represents the magnitude of impact on model output, whether it is impacting positively or negatively.

## 3. Results and discussion

### 3.1. Dataset visualization

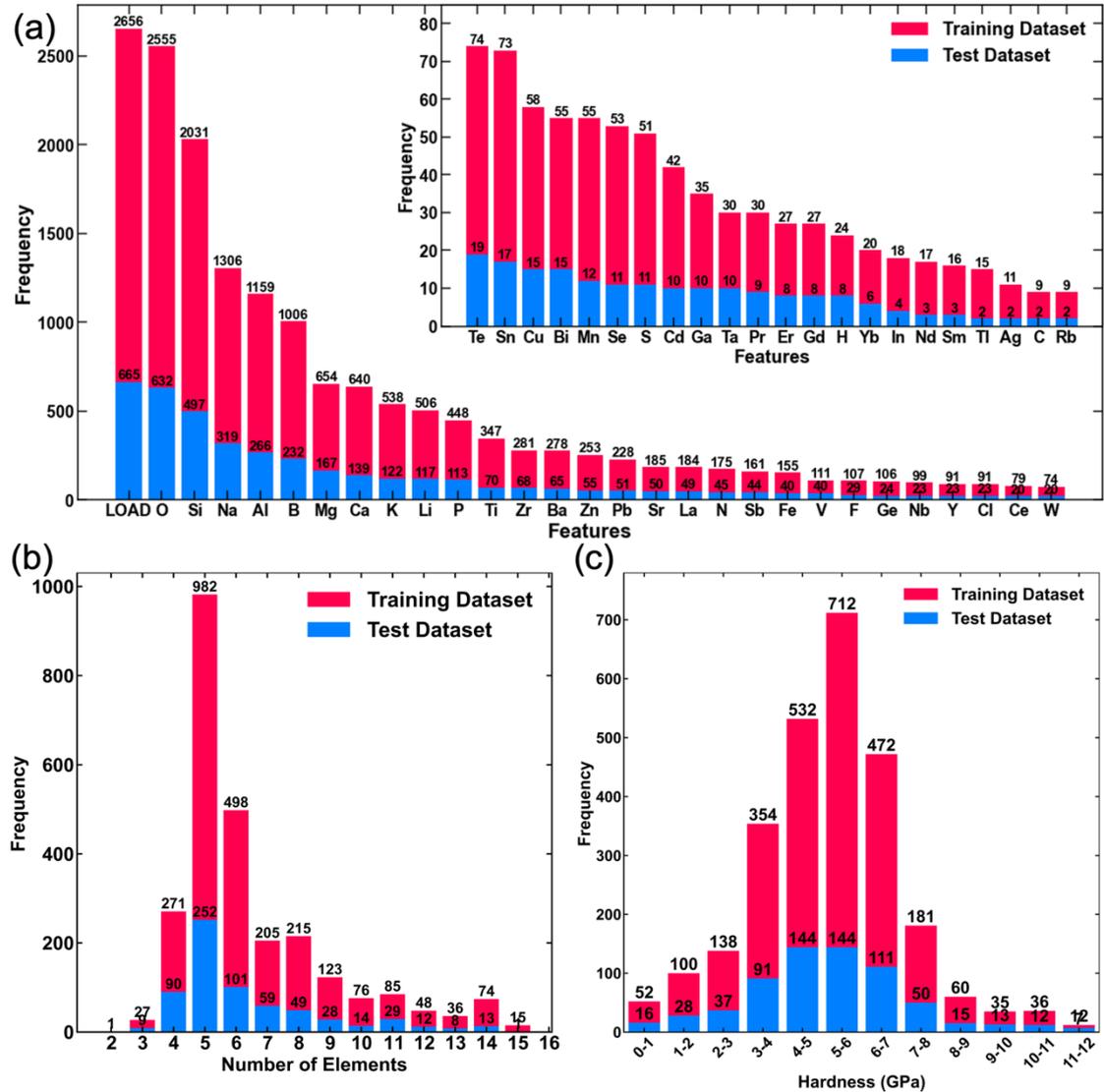

**Figure 2.** Visualizing the hardness dataset with load: (a) Number of elements and their frequency, (b) Frequency of glasses versus number of elements in their composition, (c) Distribution of hardness values in the datasets used for training and testing.

First, we analyze the hardness and load data compiled from the literature. Figure 2(a) shows the bar chart of glass compositions associated with each feature of the training and test sets. The training and test sets have 2656 and 665 glasses, respectively, of which 2555 and 632 are oxide glasses, and the others are oxynitrides, oxy-halides, or chalcogenide glasses. The inset of Fig. 2(a) shows the components present in a relatively small number of compositions. Figure 2(b) shows the distribution of the number of chemical elements present in each glass composition in the training and test sets, that is, how many compositions present are two-, three-, four-component, etc. The most frequent value is for glasses having five elements, followed by six and four. The compositions with the most components in the dataset contain five elements. Figure 2(c) shows the frequency of glass compositions in different hardness ranges. The glasses having hardness < 2 GPa are mostly phosphosilicates, and chalcogenides, while the most frequent hardness values are within 5-6 GPa. The glasses with high hardness are mostly oxynitrides with hardness values of 11-12 GPa. Thus, the dataset presented here covers a broad range of composition and hardness values for inorganic, non-metallic glasses.

## 3.2. Machine learning models

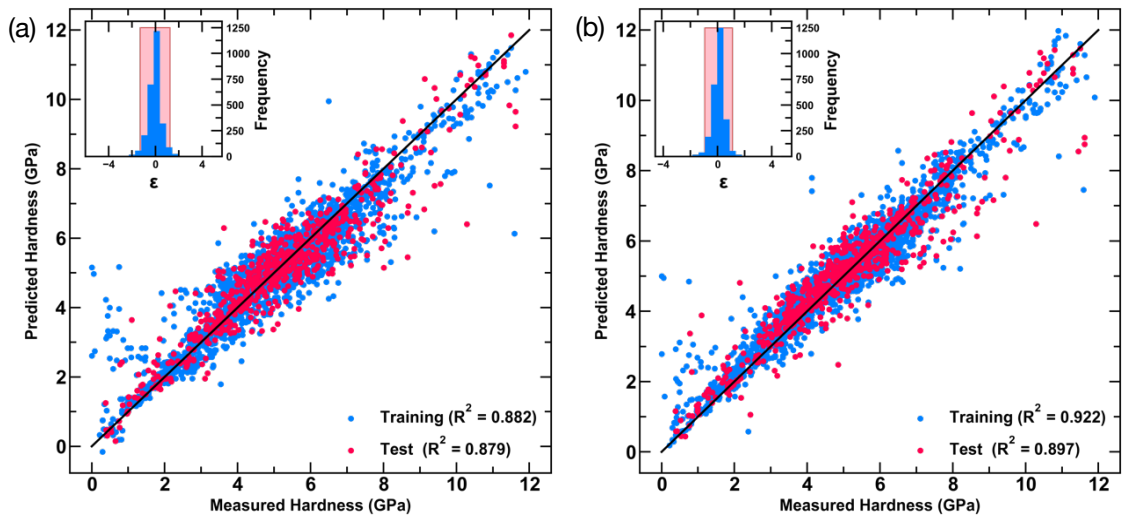

**Figure 3.** Predicted values of hardness (a) without load and (b) with the load as input features using the ML models with respect to the experimental values for both the training and test datasets. Inset shows a histogram of errors, where the shaded region represents the 95% confidence interval.

Figures 3(a) and 3(b) show predicted hardness values with and without load as input features, respectively, compared to the experimentally measured values. These models will henceforth be referred to as ML hardness with load (MLHL) and ML hardness without load (MLH). We observe that the MLHL model, which takes both composition and load as input features, performs better than the MLH model, which takes only the composition as an input feature (training and test $R^2$ scores of 0.922 and 0.897, respectively, against 0.882 and 0.879, respectively). These results exemplify the importance of the applied load for hardness measurements and predictions. The insets represent the error distribution as a probability density function. The narrow region of the red-colored shaded portion indicates the 95% confidence limit, confirming a small error for most of the values. Altogether, we observe that the model with the load as an input feature exhibits reliable hardness predictions. It should be noted that although the model predictions are relatively good, it is unclear if it can capture the load dependence on the hardness, which will be discussed in detail later.

*3.3. Symbolic reasoning informed ML*

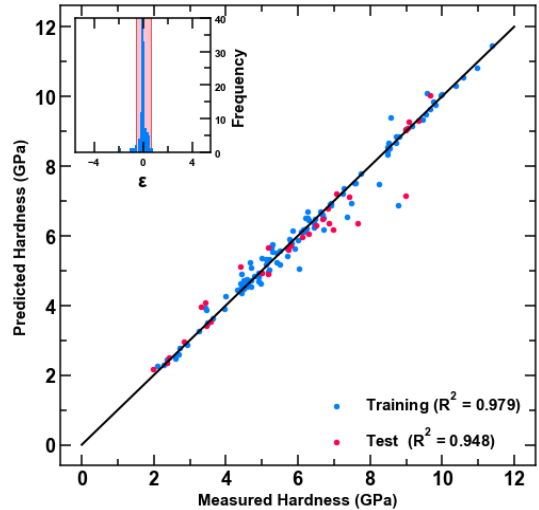

**Figure 4.** Predicted hardness values using SRIMP with respect to the experimental values for both the training and test datasets. A histogram of errors is plotted in the inset, where the shaded region represents the 95% confidence interval.

Now, we evaluate the performance of the induced SRIMP model to predict the hardness values. For the model to be trained effectively, we need at least three data points corresponding to each glass (composition, three load values, and hardness) since there are two fitting parameters, namely $C_1$ and $C_2$. Thus, the dataset employed to train the SRIMP model is significantly smaller, with ~150 glass compositions (see Supplementary materials for details). Figure 4 shows the predicted hardness values using the SRIMP approach compared to the measured values. The model performance, as given by the $R^2$ values (training and test values of 0.979 and 0.948, respectively), is excellent in predicting glass hardness.

*3.4. Indentation size effect*

To evaluate the ability of the models to generalize to unseen glass compositions and loads, we consider recent indentation experiments done on three new glass compositions synthesized in-house, that is, $25Na_2O-37.5B_2O_3-37.5SiO_2$ (NBS), $25Na_2O-75B_2O_3$ (NB), and $25Na_2O-75SiO_2$ (NS), compositions given in mol%. The hardness of these compositions was evaluated at multiple loads experimentally [41]. To quantify how different the new compositions are from the compositions in the training set, we use the Manhattan distance as a metric (see Supplementary materials). Figure 5 shows the minimum Manhattan distance for glass compositions used in the training and test set with their closest neighbor in the training dataset. We have also computed the minimum Manhattan distance for three selected glasses prepared and tested in-house: $25Na_2O-37.5B_2O_3-37.5SiO_2$ (NBS), $25Na_2O-75B_2O_3$ (NB), and $25Na_2O-75SiO_2$ (NS), respectively (Fig. 5). It is worth noting that no glass composition belonging to the NBS family exists in the training set. Most important is that Fig. 5 shows that the NBS composition is far away from those used for model training, hence it provides a test for the extrapolation ability of the SRIMP.

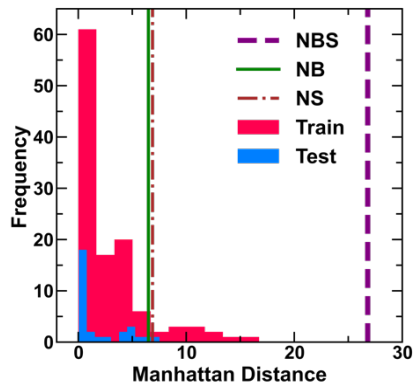

**Figure 5.** Histogram of the Manhattan distance between a glass composition and its nearest neighbor from the training set along with three selected glass compositions prepared and tested in-house, $25Na_2O–37.5B_2O_3–37.5SiO_2$ (NBS), $25Na_2O–75B_2O_3$ (NB), and $25Na_2O–75SiO_2$ (NS). The compositions are in mol%.

To evaluate the ability of the ML models to capture the ISE, the hardness of these compositions was predicted using MLHL and SRIMP models. Note that the MLH model was not used as it cannot capture the load dependency. Figures 6(a-c) show the hardness predicted by MLHL and SRIMP models, respectively. We observe that the data-driven MLHL model cannot capture the load-dependent hardness variation of the glass compositions. Even the trend of the hardness predicted by the MLHL model is not representative of the ISE in glasses. Thus, MLHL performs poorly in extrapolating beyond trained compositions and loads and does not capture the ISE.

In contrast, the SRIMP model not only correctly predicts the ISE trend of hardness but also provides values reasonably close to the experimentally obtained ones that the model has no knowledge of. Thus, the SRIMP model captures the ISE, qualitatively and quantitatively, for completely unseen compositions and loads.

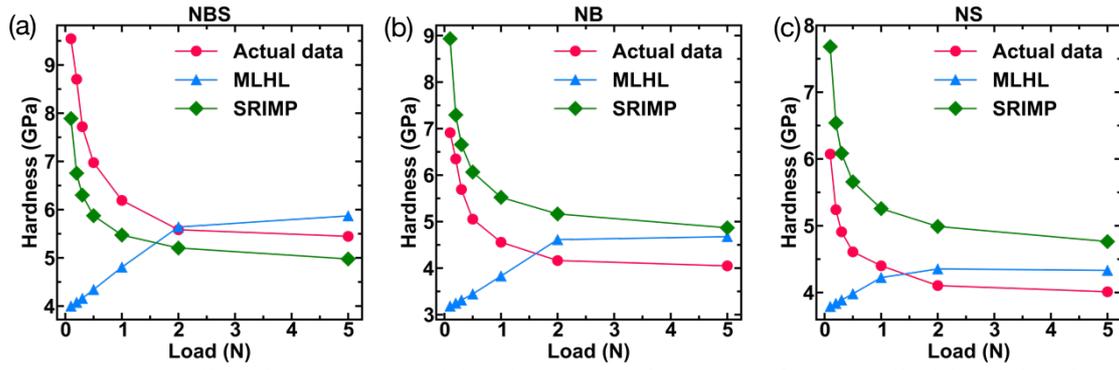

**Figure 6.** Predicted versus actual hardness as a function of the applied load for three selected glass compositions prepared and tested in-house, (a) 25Na$_2$O–37.5B$_2$O$_3$–37.5SiO$_2$ (NBS), (b) 25Na$_2$O–75B$_2$O$_3$ (NB), and (c) 25Na$_2$O–75SiO$_2$ (NS), with the compositions in mol% using MLHL, and SRIML model.

*3.5. Load-independent hardness and indentation size effect*

To interpret the formulation of the SRIMP, we now evaluate the load-dependent ($a_{\text{ISE}}$) and load-independent ($H_\infty$) components of the hardness (see Equations (4) and (5)). Figure 7(a-c) shows the actual versus predicted hardness for the NBS, NB, and NS compositions, where the pink and blue shaded regions represent the contribution of load-dependent ($a_{\text{ISE}}$) and independent hardness ($H_\infty$), respectively, toward the hardness at any value of the load. As expected from the formulation, we observe that the $H_\infty$ represents the asymptotically convergent hardness value. This value remains constant irrespective of the applied load. Thus, $H_\infty$ can be used to describe the inherent hardness of the glass. In contrast, the load-dependent $a_{\text{ISE}}$ decreases monotonically, asymptotically tending to zero with increasing applied load. Thus, the contribution of $H_\infty$ towards the hardness is significantly more than the $a_{\text{ISE}}$, except for low load values. It should be noted that both $H_\infty$ and $a_{\text{ISE}}$ are computed directly from the $C_1$ and $C_2$ terms predicted by the SRIMP model. Thus, the SRIMP formulation provides a framework to decouple the contribution of ISE in hardness and directly predict them from the glass composition.

However, due to the black-box nature of the NNs, it is not clear how each of the components in the composition controls the $C_1$ and $C_2$ terms, and in turn, $a_{ISE}$ and $H_\infty$. Therefore, we use the SHAP to interpret the effect of different elements present in the glass on the Vickers hardness.

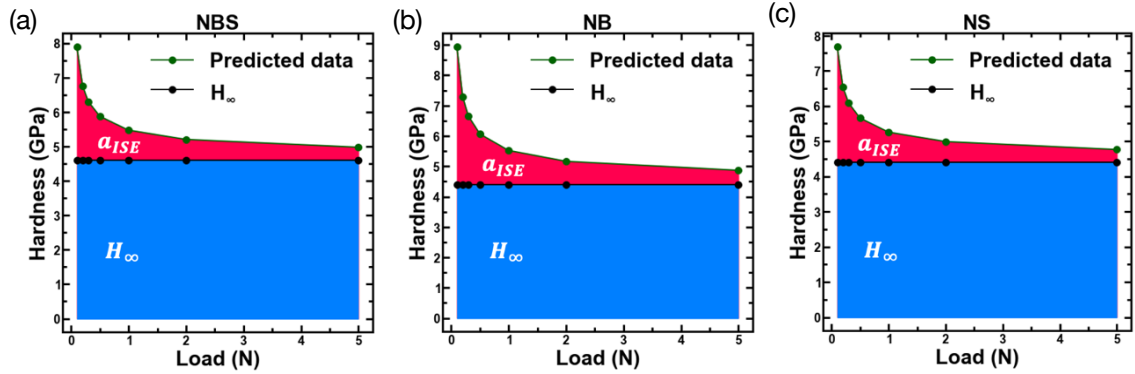

**Figure 7.** Actual versus predicted hardness as a function of applied load for (a) NBS, (b) NB, and (c) NS glass compositions using the SRIMP model.

*3.6. Model interpretation using SHAP*

SHAP is a model-agnostic post-hoc tool that can be used to interpret ML models. SHAP shows how each input feature increases or decreases the output value. Figure 8(a-b) shows the SHAP bar and violin plots explaining the MLHL predictions. We analyze the MLHL predictions in this case, as they cover a larger compositional space than the SRIMP model. Moreover, here we focus on interpreting the role of individual elements toward the hardness glasses, not the ISE. Since the MLHL model provides good predictions for the compositions considered in the dataset, it is reasonable to apply the SHAP technique to interpret the MLHL model. While the bar plots reveal the mean absolute effect of a feature on the model predictions, violin plots indicate the directionality of the feature's impact on the model predictions, whether positively or negatively.

Figure 8(a) shows that the top ten chemical elements governing the hardness of inorganic glasses are nitrogen (N), silicon (Si), sodium (Na), lithium (Li), phosphorous (P), and aluminum (Al), magnesium (Mg), boron (B), calcium (Ca), and lead (Pb). However, while some of these features impact the hardness positively, others have a negative impact. To this end, we analyze the role of these elements through the violin plots (Fig. 8(b)), where the x-axis represents the SHAP value in GPa (that is, the contribution toward hardness by a given datapoint), and color represents the normalized feature value. Among these top ten elements, while N, Si, Ca, and Mg positively affect the hardness (they increase it), Na, P, and Pb have negative effects. Other elements that increase the hardness are La, Zr, Ti, and Zn. Interestingly, Al, B, and Li have mixed effects, that is, the effect of these elements on hardness depends on the other species present in the system. Note that elements such as B and Al are well-known to have anomalous effects due to their unique structure characterized by the Loewenstein rule and the boron anomaly [9,58,59].

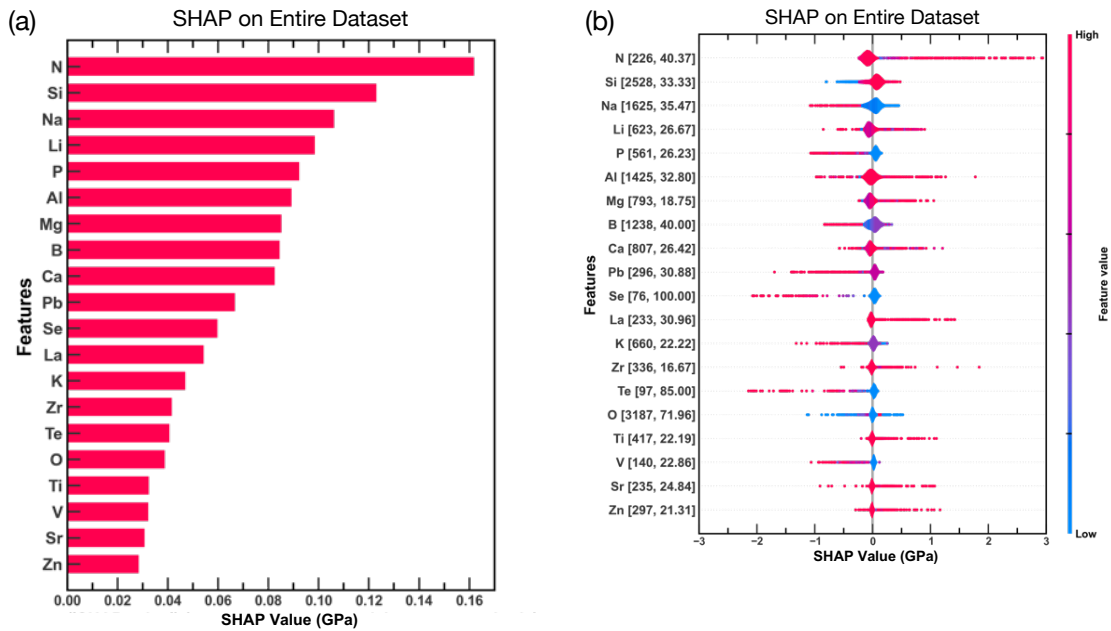

**Figure 8.** SHAP plot for the (a) absolute impact of the elements on hardness and (b) impact of the elements on hardness. In (b), the color of points represents the feature

value, that is, in terms of the mol% of the oxide components in the glasses, with red representing high contents of the element and blue representing low values in the glass composition.

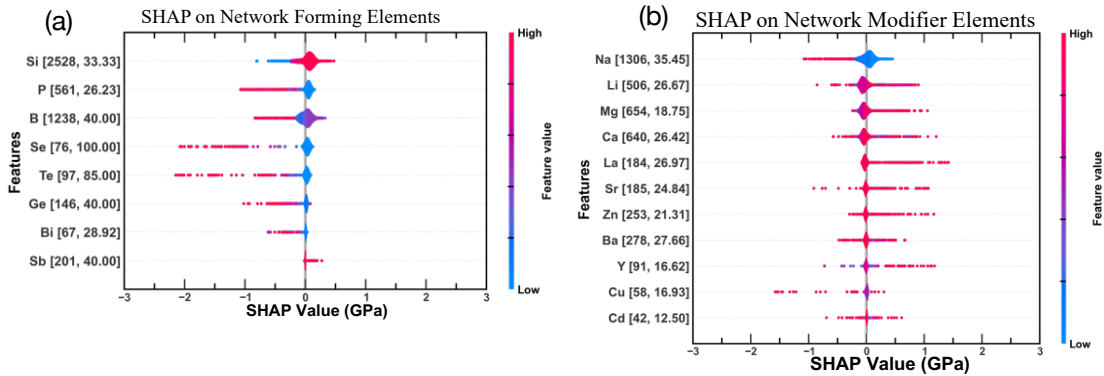

**Figure 9.** SHAP values for (a) the glass network forming elements and (b) the glass network modifying elements. The color of points represents the magnitude of the mol % of the elements.

Traditionally, glass properties are analyzed and understood in terms of the network forming and network modifying species present in the glass structure. To interpret the MLHL model similarly, we performed a SHAP analysis separately for the network forming and modifying elements (see Fig. 9). We observed that most of the network formers, except Si and B, decrease the hardness with increasing concentrations. To understand this, we can consider an existing model of glass hardness based on topological constraint theory [60]. Predicting hardness depends on the density of rigid bond constraints. That is, both the connectivity of the network and the atomic packing density govern glass hardness. For example, this explains why pure $SiO_2$ glass is relatively hard but is not among the hardest known oxide glasses. Neglecting some defects, $SiO_2$ glass has a fully polymerized structure with four bridging oxygens per tetrahedron, but this connectivity is counterbalanced by its very low packing density

(open structure) [61]. Furthermore, network formers, such as P, have a double bonded terminal oxygen, which is not considered a part of the network. Therefore, a fully polymerized phosphorous atom is effectively three-fold coordinated, providing fewer bond constraints per atom than silicon [62].

Figure 9(b) indicates that Na decreases the hardness. In contrast, other network modifiers either increase the hardness or exhibit mixed effects, that is, they exhibit different behavior depending on the other elements in the glass composition. This complex behavior can be understood as follows. While the network modifiers, such as Na, will generally tend to decrease the network connectivity by breaking bridging oxygen bonds, they can also increase the atomic packing density (e.g., in the case of silicate glasses), thus creating competing effects on glass hardness, as discussed above. Moreover, in the case of, e.g., borate-containing glasses, modifiers can also increase the connectivity of network formers by converting boron from three- to four-fold coordination. Altogether, the analysis presented here can be useful for identifying the components that, on average, affect hardness positively or negatively, thereby enabling experimentalists to design new glasses in a rational and effective fashion.

4. Conclusions

We used ML and symbolic reasoning to describe, predict, and interpret the chemical composition and load dependency of the hardness of inorganic glasses. To this end, we constructed a carefully curated dataset of glasses. We then employed a NN model for predicting the glass hardness, which exhibits good performance but is unable to capture the ISE. To address this challenge, we proposed and successfully tested a SRIMP framework combining symbolic reasoning Bernhardt's law) and ML. We demonstrate that the resulting SRIMP model can capture the ISE for completely unseen compositions and load values. Further, the framework allows the deconvolution of hardness into load-independent and load-dependent terms. The load-independent hardness proposed here can be used as an inherent material property, which along with the load-dependent term, can predict the hardness corresponding to any load. Such a framework allows for easy and rational comparison of the glass hardness measured at different loads. Finally, employing the SHAP analysis, we interpret the black-box ML models to identify the role of individual elements in governing the hardness. Such an approach enables the accelerated design of glasses with targeted hardness.

**ACKNOWLEDGMENTS:**

N.M.A.K. acknowledges the financial support for this research provided by the Department of Science and Technology, India, under the INSPIRE faculty scheme (DST/INSPIRE/04/2016/002774) BRNS YSRA (53/20/01/2021-BRNS) and DST SERB Early Career Award (ECR/2018/002228) award by the Government of India. The authors thank the IIT Delhi HPC facility for providing computational and storage resources. E.D.Z. acknowledges the São Paulo Research Foundation (FAPESP Cepid grant 2013/007793-6) for generous funding since 2013. M.M.S. acknowledges funding by the European Union (ERC, NewGLASS, 101044664). D.R.C acknowledges the INCT – Materials Informatics project financial support. Views and opinions expressed are, however, those of the authors only and do not necessarily reflect those of the European Union or the European Research Council. Neither the European Union nor the granting authority can be held responsible for them.